\documentclass[aps,prb, twocolumn, preprintnumbers,amsmath,amssymb,superscriptaddress, floatfix]{revtex4}
\usepackage{graphicx, bm}
\usepackage{latexsym}
\usepackage{amsmath}
\usepackage{graphics}
\usepackage{amssymb}
\usepackage{layout}
\usepackage{verbatim}
\usepackage{amsfonts,epsfig}

\begin{document}
\title{Quantum decoherence of a charge qubit in a spin-fermion model}
\author{Roman M. Lutchyn}
%\email{lutchyn@umd.edu}
\affiliation{Condensed Matter Theory Center, Department of Physics, University of Maryland, College Park, MD 20742-4111, USA}
\affiliation{Joint Quantum Institute, Department of Physics, University of Maryland, College Park, MD 20742-4111, USA}
\author{{\L}ukasz Cywi{\'n}ski}
%\email{lcyw@umd.edu}
\affiliation{Condensed Matter Theory Center, Department of Physics, University of Maryland, College Park, MD 20742-4111, USA}
\author{Cody P. Nave}
\affiliation{Condensed Matter Theory Center, Department of Physics, University of Maryland, College Park, MD 20742-4111, USA}
\author{S. Das Sarma}
\affiliation{Condensed Matter Theory Center, Department of Physics, University of Maryland, College Park, MD 20742-4111, USA}
\affiliation{Joint Quantum Institute, Department of Physics, University of Maryland, College Park, MD 20742-4111, USA}
\date{
%resub PRB v4 - edited by RL and LC,
 \today}

\begin{abstract}

We consider quantum decoherence in solid-state systems by studying the transverse dynamics of a single qubit interacting with a fermionic bath and driven by external pulses. Our interest is in investigating the extent to which the lost coherence can be restored by the application of external pulses to the qubit.
We show that the qubit
evolution under various pulse sequences can be mapped onto Keldysh
path integrals. This approach allows a simple diagrammatic treatment
of different bath excitation processes contributing to qubit
decoherence. We apply this theory to the evolution of the qubit
coupled to the Andreev fluctuator bath in the context of widely studied superconducting qubits. We show that charge fluctuations within the Andreev-fluctuator model lead to a $1/f$ noise spectrum with a characteristic temperature depedence. We discuss the strategy for suppression of decoherence by the application of higher-order (beyond spin echo) pulse sequences.

\end{abstract}

\maketitle

\section{Introduction}
The loss of coherence of a  quantum  two-level system (quantum bit) is caused by its unavoidable coupling to the surrounding environment. For solid-state qubits, the decoherence process can be quite fast due to coupling to a large number of internal degrees of freedom. Our understanding of quantum decoherence and methods for its suppression in a realistic solid-state environment is mainly confined to the cases of a qubit interacting with bosonic\cite{Leggett_RMP87, Uhrig_PRL07, Makhlin_PRL04} and nuclear spin baths,\cite{Khaetskii_PRB03, Witzel_PRB06, Saikin_PRB07} the so-called (and extensively studied) spin-boson and spin-bath models, respectively. A less well understood, but very relevant case for solid-state quantum architectures is that of a qubit coupled to a fermionic bath,\cite{Chang_PRB85,Paladino_PRL02,Grishin_PRB05,Grishin_06, deSousa_PRL05, Segal_PRB07, Abel_08} which dramatically differs from the previous examples.
In this paper we  study quantum decoherence in the context of a superconducting charge qubit\cite{Makhlin_RMP01,Nakamura_PRL02,Vion_Science02,Ithier_PRB05,Astafiev_PRL04,Astafiev_PRL06,Schreier_PRB08} interacting with the non-trivial bath of Andreev fluctuators.\cite{Faoro_PRL05, Kozub_PRL06, Faoro_Ioffe_PRL06} This problem is a paradigmatic spin-fermion decoherence problem and applies to many situations involving the quantum coupling of a qubit (``spin'') to a general fermionic environment.
Using a many-body  Keldysh path integral approach,\cite{Levitov_review_Keldysh, Nazarov_EurJB03, Kamenev_04,Grishin_PRB05,Grishin_06} we obtain a quantum-mechanical description of the qubit evolution under pulse sequences aimed at prolonging the coherence of the system. The simplest case of decoherence under pulses is the spin echo dephasing experiment, which has been shown to extend the coherence time of solid state (superconducting) qubits,\cite{Nakamura_PRL02,Ithier_PRB05, Wallraff_2007} by essentially eliminating the quasi-static shifts of qubit energy splitting (inhomogeneous broadening) due to the slow environmental fluctuations.
However, sequences involving more pulses, for example,~CPMG~\cite{Haeberlen} and Uhrig's \cite{Uhrig_PRL07,Uhrig_08} sequences, are expected to lead to a further increase of the coherence time.\cite{Viola_PRA98,Viola_JMO04,Falci_PRA04,Khodjasteh_PRL05,Khodjasteh_PRA07,Faoro_PRL04,Uhrig_PRL07,Lee_PRL08,Cywinski_PRB08,Uhrig_08}

In this paper, we consider an experimentally relevant example - a superconducting qubit coupled to fluctuating background charges,\cite{Astafiev_PRL04, Astafiev_PRL06} e.g.~electrons residing on Anderson-impurity sites. Due to a large on-site Coulomb repulsion forbidding double occupancy, this example represents a non-trivial interacting bath. The dynamics of the charge fluctuations on the impurity sites is determined by the hybridization of impurity levels with the quasiparticle band of the superconductor. To the lowest order in tunneling at the superconductor/insulator interface, the hybridization of the impurity levels can be described by a correlated tunneling events of two electrons with opposite spin to/from the superconductor.  We show that in the small background-charge density limit, these fluctuations lead to a $1/f$ spectral density of noise.  Using these results, we finally obtain the quantum-mechanical description of the qubit evolution driven by external pulses, and discuss optimal strategy for the suppression of the decoherence with designed composite pulse sequences.

The paper is organized as follows: In Sec.~\ref{general} we provide a general derivation of the qubit evolution with pulses and map the calculation of decoherence function onto Keldysh path integral formalism. In Secs.~\ref{AndreevBath} and~\ref{Andreevspectral}  we introduce Andreev fluctuator bath and derive spectral density of noise for this model. Finally, in Sec.~\ref{sec:F} we discuss the influence of pulse sequences on the qubit decoherence.

\section{General theory for qubit evolution }\label{general}
The transverse dynamics of a
qubit interacting with its environment is determined by the following Hamiltonian
\begin{equation}
\hat{H} = \frac{E}{2}\hat{\sigma}_{z} + \hat{\sigma}_{z}\hat{V}+
\hat{H}_{\rm B}\,\, . \label{eq:H}
\end{equation}
Here the environment is represented by a fermionic bath $\hat{H}_{\rm B}$, and the qubit is
coupled to the environment through the density fluctuation operator:
\begin{eqnarray}
\hat{V}&=&\sum_{l\sigma}v_l (c^\dag_{l\sigma} c_{l\sigma}-\langle n_{l\sigma}\rangle ) \,\,  . \label{eq:V}
\end{eqnarray}
This model corresponds precisely to the coupling of
a superconducting charge qubit to the density fluctuations on the
impurities in the substrate. Here $c_{l\sigma}$ and $c^\dag_{l\sigma}$ are the fermionic
annihilation and creation operators at $l$-th site with spin $\sigma$, and $v_l$ and  $\langle n_{l\sigma} \rangle$ are,
respectively, the strength of the coupling and average occupation
of $l$-th impurity, i.e.~$\langle n_{l\sigma} \rangle=\langle c_{l\sigma}^\dag c_{l\sigma}\rangle$. Equations~(\ref{eq:H}) and (\ref{eq:V}) define our spin-fermion model.

We study the evolution of the qubit in contact with a fermionic bath
assuming the qubit energy relaxation time $T_1$ to be much longer
than the quantum dephasing time $T_2$ (thus only $\hat{\sigma}_{z}$ coupling is present in the Hamiltonian).
Qubit decoherence under the influence
of the environment is given by the off-diagonal matrix elements of
the qubit's reduced density matrix, and for the free evolution of the qubit we get  ($\hbar=1$)
\begin{equation}\label{rhopm}
\rho_{+-}(t) =   \langle + | \text{Tr}_{B} \{ \hat{\rho}(t) \} |-\rangle   =  \rho_{+-}(0)e^{-i E t} W(t).
\end{equation}
In the above $\hat{\rho}(t)$ is the density matrix of the whole system (qubit$+$bath), which is assumed factorizable at $t \! = \! 0$, $\text{Tr}_{B}\{...\}$ is the trace with respect to the bath degrees of freedom,
and $W(t)$ is the decoherence function defined as
 \begin{equation}
W(t)=\left \langle e^{i(\hat H_{\text{B}}-\hat V)t}e^{-i(\hat H_{\text{B}}+\hat V)t} \right \rangle
 \end{equation}
with the brackets representing the thermal average with respect to the bath
Hamiltonian $\hat{H}_{\text{B}}$, i.e. $\langle ... \rangle \! = \! \text{Tr}_{B} \{ \hat{\rho}_{B} ... \}$. The time $t$ always refers here to the total time of the evolution.

In addition to the free evolution of the qubit (free induction decay), one is often interested in the dynamics of the system subject to external $\pi$-pulses\cite{Viola_PRA98,Viola_JMO04,Falci_PRA04,Faoro_PRL04,Khodjasteh_PRL05,Khodjasteh_PRA07,Uhrig_PRL07,Uhrig_08,Lee_PRL08,Cywinski_PRB08}  which could, in principle, prolong or restore quantum coherence. The $\pi$-pulses considered here correspond to rotations of the qubit's Bloch vector by angle $\pi$ about, e.g., the $\hat{x}$ axis, and are short enough for the bath dynamics during the pulse duration to be negligible. Then, the evolution operator for qubit and bath is given by
\begin{equation}
\hat{U}^{(n)}(t) = (-i)^{n} \, e^{-i\hat{H}\tau_{n+1}} \hat{\sigma}_{x} e^{-i\hat{H}\tau_{n}} \! ... \! \hat{\sigma}_{x} e^{-i\hat{H}\tau_{1}}
\end{equation}
with $n$ and $\tau_{i}$ being the number of applied pulses and time delays between the pulses,respectively, and the total evolution time $t \! = \! \sum_{i=1}^{n+1} \tau_{i}$. One can see that the well-known Hahn spin echo (SE) sequence, for example, corresponds to a single pulse with $\tau_{1} \! = \! \tau_{2} \! = \! t/2$.

Using the fact that in the ``pure dephasing'' case under consideration, the qubit states $|\pm\rangle$ are the eigenstates of the Hamiltonian (\ref{eq:H}), we can write the decoherence function under the influence of pulses as
\begin{equation}
 W_{n}(t)\! =\!  \left \langle \left(\hat{U}^{(n)}_{-}(t)\right)^{\dagger} \hat{U}^{(n)}_{+}(t) \right \rangle  \label{eq:Wn}
\end{equation}
with the evolution operators $\hat{U}^{(n)}_{\pm}(t)$ given by
\begin{align}
\!\!\!\!\!\!\hat{U}^{(n)}_{+}(t)\!&\!=\!e^{-i(\hat{H}_{\text{B}}+\hat{V})\tau_{n+1}}e^{-i(\hat{H}_{\text{B}}-\hat{V})\tau_{n}} ... e^{-i(\hat{H}_{\text{B}}+ p \hat{V})\tau_{1}} \nonumber \,\, , \\ \label{eq:Upm} \\
\!\!\!\!\!\!\hat{U}^{(n)}_{-}(t)\!&\!=\!e^{-i(\hat{H}_{\text{B}} -\hat{V})\tau_{n+1}}e^{-i(\hat{H}_{\text{B}} +\hat{V})\tau_{n}} ... e^{-i(\hat{H}_{\text{B}} - p\hat{V})\tau_{1}} \,\, , \nonumber
\end{align}
where  $p \! = \! (-1)^{n}$ is the parity of the sequence.
Then, the off-diagonal elements of the qubit density matrix are given by
\begin{align}
\rho_{+-}(t) = \rho_{p,-p}(0) \,  e^{-i p E (\tau_{1} - \tau_{2} + ... + p\tau_{n+1})}  W_{n}(t).\\ \nonumber \,\,
\end{align}
Here the phase factor is zero for all balanced sequences (for which the total times of evolution due to $\hat{H} \! + \! \hat{V}$ and $\hat{H}\! - \! \hat{V}$ are the same in Eq.~(\ref{eq:Upm})).
The evolution of the qubit under SE sequence, for example, acquires a simple form
\begin{equation}
\rho_{+-}^{SE}(t)  \!=\!  \rho_{-+}(0)  \left \langle e^{i\hat{H}_{+}\frac{t}{2}}  e^{i\hat{H}_{-}\frac{t}{2}}  e^{-i\hat{H}_{+}\frac{t}{2}}  e^{-i\hat{H}_{-}\frac{t}{2}} \right \rangle \,\,
\end{equation}
with $\hat{H}_{\pm} \! = \! \hat{H}_{B} \pm \hat{V}$.

\begin{figure}[b]
\centering
\includegraphics[width=1.0\linewidth]{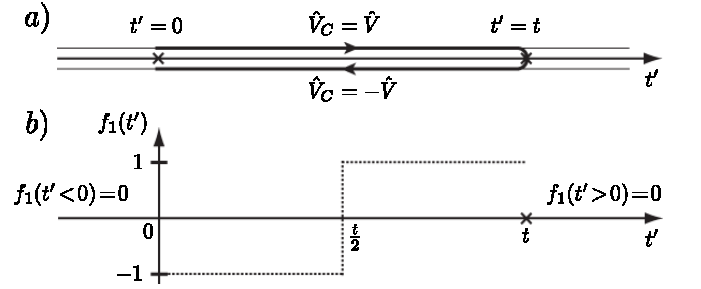}
  \caption{ a) Dependence of $\hat{V}_C(t)$ on time along the Keldysh contour. b)
 The plot of the function $f_n(t')$ for the Spin Echo sequence
($n=1$).} \label{fig:contour}
\end{figure}

Decoherence under pulses has been analyzed with methods specific to the spin-boson model\cite{Uhrig_PRL07,Uhrig_08} and the spin bath model,\cite{Witzel_PRB06, Saikin_PRB07} or using operator algebra.\cite{Khodjasteh_PRL05,Khodjasteh_PRA07,Lee_PRL08}  The latter approach, although very general, does not allow for transparent understanding of physics of the bath.
However, the evaluation of $W_{n}(t)$ defined in Eq.~(\ref{eq:Wn}) can be mapped onto the evolution on the Keldysh contour,\cite{Rammer} putting the calculation of decoherence into the framework of many-body theory. Similar formalism has been used to study full counting statistics of a general quantum mechanical variable and has  proved to be quite convenient.\cite{Levitov_review_Keldysh, Nazarov_EurJB03}
The evolution operators $\hat{U}^{(n)}_{\pm}$  can be written as
\begin{equation}
\hat{U}^{(n)}_{\pm}(t) = \mathcal{T} \exp \left [ -i \int_{0}^{t} (\hat{H}_{B} \pm f_{n}(t') \hat{V}) dt' \right ] \,\, ,
\end{equation}
where $\mathcal{T}$ is the time ordering operator. The function $f_{n}(t')$ encodes a particular sequence, and is defined as
\begin{equation}
f_{n}(t') = p \sum_{k=0}^{n} (-1)^{k} \Theta(t_{k+1}-t')\Theta(t'-t_{k}),
\end{equation}
where $\Theta(t')$ is the Heaviside step function, $t_{k}$ with $k \! = \! 1..n$ are the times at which the pulses are applied, $t_{0}\!= \! 0$, and $t_{n+1} \! = \! t$.
Thus, the product of operators inside the average in Eq.~(\ref{eq:Wn}) corresponds to (reading from left to right) the time-ordered evolution from $0$ to $t$ (with $+\hat{V}$ coupling), followed by the time anti-ordered evolution from $t$ to $0$ (with $-\hat{V}$ coupling). We can then introduce the Keldysh contour $C$ (see Fig.~\ref{fig:contour}a) together with the notion of contour-ordering of operators.\cite{Rammer,Kamenev_04} The qubit-bath coupling takes then two opposite signs on the upper/lower branch of the contour: $\hat{V}_{C} \! = \! \pm \hat{V}$. While $f_{n}(t')$ is non-zero only for $t'\in [0,t]$, we can extend the limits of time integration on both branches to $[-\infty,\infty ]$. The evolution from $t' \! = \! -\infty$ allows one to include the adiabatically turned-on interactions in $\hat{H}_{B}$ (see, for example, Ref.~[\onlinecite{Kamenev_04}]), paving the way to the treatment of decoherence in an interacting fermionic bath. The final result is most compactly written as a functional integral with the Grassmann fields $\bar{\psi}_l$ and
$\psi_l$ defined on the Keldysh contour:\cite{Kamenev_04,
Rammer,Grishin_PRB05,Grishin_06}
\begin{widetext}
\begin{align}
\!\!\!\!W_{n}(t)\!\!=\!\!
\left \langle \! \mathcal{T_{C}} \exp\!\!\left(\!-i\!\!\int_{C}dt' \! \left[\hat{H}_B\!+\!\hat{V}_{C} f_n(t')\right]\! \right)  \! \right \rangle\!\!=\!\!
\frac{1}{Z_B}\!\int
\!\!\!   \mathcal{D}\bar{\psi}_l\mathcal{D}\psi_l \exp \!\! \left( \!\! i
S_B\!\left[\bar{\psi}, \psi\right]\!-\!i\!\int\limits_{C}dt'\!
\sum\limits_{l\sigma} \!v_l(t')\!f_{n} (t')\!\!\left[\bar{\psi}_{l\sigma}(t')\psi_{l\sigma}(t')\!-\!\langle n_{l\sigma}\rangle\!\right]
\!\! \right)\!\!, \label{funcint}
\end{align}
\end{widetext}
where the integration is performed on the contour $C$ shown in
Fig.~\ref{fig:contour}a, $v_{l}(t') \! = \! \pm v_{l}$ on the
upper/lower branch of the contour, and the normalization constant is
defined as the functional integral with $\hat V \! = \! 0$. The bath
action $S_{B} \! = \! S_{0} + S_{int}$, and the functional
integration with non-interacting $S_{0}$  corresponds to averaging
over an equilibrium noninteracting density matrix at $t' \! = \!
-\infty$. This formulation of the decoherence problem enables one to
use techniques and approximations developed in many-body theory. It
also allows for a transparent treatment of the physics of the bath
while simply encoding the driving of the qubit in a single function
of time $f_{n}(t')$.

\section{Andreev fluctuator bath}\label{AndreevBath}
\begin{figure}
\centering
\includegraphics[width=0.7\linewidth]{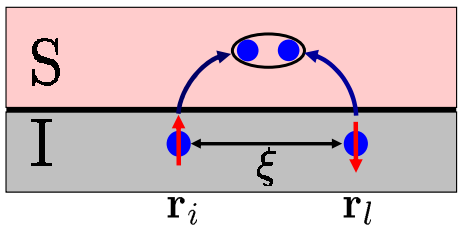}
  \caption{(color online). Correlated tunneling of two electrons with opposite spins from the impurity sites in the insulator into the superconductor. An electron from the $i$-th impurity with energy below the gap $\Delta$ tunnels into superconductor, propagates over distances of the order of coherence length $\xi$ and recombines with another electron with opposite spin from $j$-th site into a Cooper-pair. The amplitude for such Andreev process decays exponentially with distance between the impurity sites $A_{lj}\propto \exp(-|\bm r_l-\bm r_j|/\pi\xi)$, see Eq.~(\ref{eqn:amplitude}). } \label{fig:tunneling}
\end{figure}

In order to evaluate the functional integral~(\ref{funcint}), one
needs to specify the bath Hamiltonian. Here, as an example, we
consider a non-trivial bath of Andreev
fluctuators,\cite{Faoro_PRL05, Kozub_PRL06, Faoro_Ioffe_PRL06} which
describes the fluctuations of the occupation of impurities close to
insulator/superconductor interface due to Andreev processes. This
model takes into account coherent processes of creation
(destruction) of the Cooper pair in the superconductor by correlated
tunneling of two electrons from (to) different impurity sites in the
insulator,\cite{Larkin, Kozub_PRL06} see also Fig.~\ref{fig:tunneling}.
In the limit when the superconducting gap energy $\Delta$ is the
largest relevant energy scale in the problem ($T, E, \varepsilon_j,
\ll \Delta$), the effective Hamiltonian for the Andreev fluctuator
bath, after integrating out superconducting degrees of freedom, is
given by
\begin{eqnarray}\label{eqn:HB}
\!\hat{H}_{\rm B}\!\!=\!\!\sum_{l\sigma}\varepsilon_l c^\dag_{l\sigma} c_{l\sigma} \!+\!U \sum_{l}\hat{n}_{l\uparrow}\hat{n}_{l\downarrow}\!+\!\sum_{l \neq j}\!\left[A^*_{lj}c^\dag_{l\uparrow} c^\dag_{j\downarrow}\!+\! \mathrm{ H.\, c.} \right]\!\!.
\end{eqnarray}
Here, $\varepsilon_l$ and $U$ are the energy of a localized electron
on $l$-th impurity (measured with respect to the Fermi energy
$\varepsilon_F$ of the conduction electrons) and repulsive on-site
interaction (assumed to be large enough to prevent double occupation
of the sites), respectively. The matrix elements $A_{lj}$, in the
limit of low transparency barrier between the insulator and
superconductor, are given by
\begin{eqnarray}\label{eqn:amplitude}
A_{lj}\approx A_0
\frac{\sin(p_F|\textbf{r}_l-\textbf{r}_{j}|)}{p_F|\textbf{r}_l-\textbf{r}_{j}|}
\,e^{-|\textbf{r}_l-\textbf{r}_{j}|/\pi \xi}\,.
\end{eqnarray}
Here $p_F$ is the Fermi momentum, $\xi$ is the
coherence length in a clean superconductor. The amplitude $A_0=2\pi^2 d^2 a N(0) T_0^2$ is
determined by the tunneling matrix element between the insulator and
superconductor $T_0$, the normal density of states in the metal
$N(0) =mp_F/\pi^2$, the localization length under the barrier $d$
and the size of the impurity wavefunction $a$.\cite{Kozub_PRL06}

\begin{figure}
\centering
\includegraphics[width=0.99\linewidth]{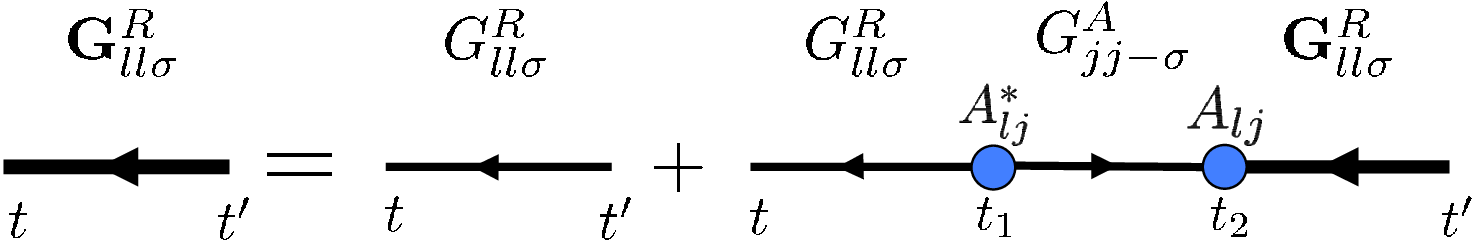}
  \caption{(color online). Dyson's equation for the retarded Green's function $\mathbf G^R$ in the Born approximation. Here we have adopted
the convention of Ref.~[\onlinecite{ Rammer}]. The advanced and Keldysh
Green's functions are obtained analogously resulting in
Eqs.~(\ref{eqn:sigma_A})-(\ref{eqn:sigma_K}). } \label{fig:Dyson_eq}
\end{figure}

Given the Hamiltonian~(\ref{eqn:HB}), the action for the bath on the
Keldysh contour can be written as
\begin{align}\label{eqn:SB}
\!\!\!S_B\left[\bar{\psi},\psi\right]&\!\!=\!\!\sum_{lj}\int
\limits_{C}dt' \sum_{\sigma} \delta_{lj} \bar{\psi}_{l\sigma}(t')(i\partial_{t'}\!-\!\varepsilon_{l}\!-\!U \langle n_{l,-\sigma}\rangle)\psi_{l\sigma}(t')  \nonumber\\
&\!+\! A^*_{lj}\bar{\psi}_{l\uparrow}(t') \bar{\psi}_{j\downarrow}(t')+A_{lj}
\psi_{j\downarrow}(t')\psi_{l\uparrow}(t') \,\, .
\end{align}
Here we used the mean-field approximation for the Anderson impurity
model assuming that the Kondo temperature $T_K$ is smaller than the
superconducting gap $\Delta$, which is reasonable in the situation at hand when the impurities are located in the substrate and the tunneling matrix element $T_{0}$ coupling them to the states in the superconductor is small. 
The
occupation probabilities $\langle n_{l\sigma} \rangle$ are obtained
self-consistently using
\begin{align}
\langle n_{l\sigma} \rangle =\int \frac{d\omega}{2\pi} \, n_F(\omega) [G_{ll\sigma}^A(\omega)-G_{ll\sigma}^R(\omega)],
\end{align}
see Ref.~[\onlinecite{Bruus}] for more details. Performing
a Keldysh rotation,\cite{Kamenev_04, Rammer} one can calculate the
full Green's function ${\bf \hat{G}}_{ll\sigma}(t,t')$ for the bath (see
Fig.~\ref{fig:Dyson_eq})
\begin{eqnarray}
{\bf \hat{G}}^{-1}_{ll\sigma}(t,t')={ \hat{G}}^{-1}_{ll\sigma}(t,t')-\hat{\Sigma}_{ll\sigma}(t,t').
\end{eqnarray}
Here ${ \hat{G}}^{-1}_{ll\sigma}(t,t')$ is the bare Green's
function, see Eq.~(\ref{eqn:SB}), and the self energy
$\hat{\Sigma}_{ll\sigma}(t,t')$ is calculated to second order in
$A_{ij}$ giving the components of the self-energy matrix
\begin{eqnarray}
 \Sigma_{ll\sigma}^{A/R}(t,t')&=&\sum_{j\neq l}  |A_{lj}|^2{G}_{jj,-\sigma}^{R/A}(t',t),\label{eqn:sigma_A}\\
 \Sigma_{ll\sigma}^K(t,t')&=&\sum_{j\neq l}  |A_{lj}|^2{G}_{jj,-\sigma}^K(t',t).\label{eqn:sigma_K}
\end{eqnarray}
In Eqs.~(\ref{eqn:sigma_A}) and ~(\ref{eqn:sigma_K}) we have
neglected the off-diagonal terms in the impurity indices, i.e.
$\Sigma_{lj\sigma}\approx\delta_{lj}\Sigma_{ll\sigma}$. Since the
amplitude $A_{lj}$ oscillates on the length scale of $p_F^{-1}$, the
contribution of these off-diagonal terms to the self-energy is
small.

Using the above results, the action for the bath can be written in
terms of the full Green's function ${\bf \hat{G}}_{ll\sigma}(t,t')$.
Then, the decoherence function becomes
 \begin{widetext}
\begin{align}\label{funcint2}
W_{n}(t)&\!\equiv\! \exp \left[-\chi_{n}(t)\right]\!=\!\frac{1}{Z_B}\!\int
\mathcal{D}\bar{\psi}\mathcal{D}\psi \\
&\times \exp\!\left[i \sum_{l\sigma} \int \limits_{-\infty}^{\infty} dt_1 \!\! \int \limits_{-\infty}^{\infty}   dt_2 \left( \sum_{a,b=1}^{2} \bar{\psi}_{l\sigma}^{(a)}(t_1)
\left[ {\bf \hat{G}}^{-1}_{ll\sigma}(t_1,t_2) \right]_{ab}
\psi^{(b)}_{l\sigma}(t_2)\!-\!2\delta(t_1\!-\!t_2)v_l f^{(n)}(t_1)\left[
\rho_{l\sigma}(t_1)\!-\! \langle n_{l\sigma} \rangle \right]\right)
\right].\nonumber
\end{align}
\end{widetext}
Here, $\rho_{l\sigma}(t)$ corresponds to the fermion density operator
$
\rho_{l\sigma}(t)=\frac12 \left[\bar{\psi}^{(1)}_{l\sigma}(t)\psi^{(2)}_{l\sigma}(t)\!+\!\bar{\psi}^{(2)}_{l\sigma}(t)\psi^{(1)}_{l\sigma}(t)\right]\!
 $;  the fields $\psi^{(1)}(t)$ and $\psi^{(2)}(t)$ are given by the
 appropriate superposition of the fermionic fields on the upper and
lower parts of the Keldysh contour, see Ref.~[\onlinecite{Kamenev_04}].
After performing the functional integral over the fermionic fields and
expanding to second order in $v_l$, one finds
\begin{eqnarray}
\!\!\chi_n(t)\!&\!\!=\!\!&\!\!\sum_{l\sigma}\frac{v_l^2}{2}\!\!\!\int_{0}^{t}\!\!\int_{0}^{t}\!\!dt_1dt_2 f_n(t_1\!) f_{n}(t_2\!)
\!\left[{\bf G}_{ll\sigma}^A(t_1,\!t_2){\bf G}_{ll\sigma}^R(t_2,\!t_1) \right. \nonumber \\ \nonumber\\
\!\!&\!\!+\!\!&\!\! \left. {\bf G}_{ll\sigma}^R(t_1,t_2){\bf G}_{ll\sigma}^A(t_2,t_1)\!+\!{\bf G}_{ll\sigma}^K(t_1,t_2){\bf
G}_{ll\sigma}^K(t_2,t_1)\!\right]\!\!. \label{chi1}
\end{eqnarray}
Equation~(\ref{chi1}) holds whenever the short-time expansion is
valid. The long-time asymptote can be obtained by resumming
the whole series.\cite{Grishin_PRB05,Grishin_06}

By introducing the Fourier transform of the Green's functions, Eq.~(\ref{chi1}) can be formally recast as
\begin{equation} \label{eqn:chi}
\chi_n(t)=\int_{-\infty}^{\infty} \frac{d\omega}{2\pi} \,
\frac{F_{n}(\omega t)}{\omega^2} S_Q(\omega).
\end{equation}
Here $F_{n}(\omega t)=\omega^2 |f_n(\omega)|^2/2$ is a
sequence-specific filter function, the role of which we discuss in
Sec.~\ref{sec:F}. Thus, to second order in $v_i$ we obtain
$\chi_{n}(t)$ having the same structure as in the case of a qubit
coupled to the spin-boson bath or classical
noise,\cite{Martinis_PRB03,Uhrig_PRL07,Uhrig_08,Cywinski_PRB08} i.e.~$\chi_{n}(t)$ is the integral of the product of the
environment-specific spectral density of noise $S_{Q}(\omega)$ and sequence-specific filter function
$F_{n}(\omega t)$. The spectral density of quantum noise
$S_Q(\omega)$ in the spin-fermion problem is given by
\begin{widetext}
\begin{align}\label{eqn:noise}
\!\!S_Q(\omega)\!\!=\!\!\sum_{l\sigma}v_l^2 \int
_{-\infty}^{\infty} \frac{d\Omega}{2\pi} \left[{\bf G}_{ll\sigma}^A\!\left(\Omega\!+\!\frac{\omega}{2}\right){\bf G}_{ll\sigma}^R\left(\Omega\!-\!\frac{\omega}{2}\right)\!+\!{\bf G}_{ll\sigma}^R\left(\Omega\!+\!\frac{\omega}{2}\right){\bf G}_{ll\sigma}^A\left(\Omega\!-\!\frac{\omega}{2}\right)\!+\!{\bf G}_{ll\sigma}^K\left(\Omega\!+\!\frac{\omega}{2}\right){\bf G}_{ll\sigma}^K\!\left(\Omega\!-\!\frac{\omega}{2}\right)\right]\!.  
\end{align}
\end{widetext}
In the frequency domain, the full Green's functions are
\begin{align}
&{\bf G}_{ll\sigma}^{A/R}(\omega)=\frac{1}{\omega-\varepsilon_l-U \langle n_{l,-\sigma} \rangle-\Sigma^{A/R}_{ll\sigma}(\omega)},\nonumber\\
&{\bf G}_{ll\sigma}^{K}(\omega)=\tanh \left(\frac{ \omega}{2T} \right)\left[{\bf G}_{ll\sigma}^{R}(\omega)-{\bf G}_{ll\sigma}^{A}(\omega)\right],
\end{align}
where the self energy $\Sigma_{ll\sigma}^{A/R}(\omega)$ is defined as
\begin{eqnarray}\label{Sigma_freq}
\Sigma_{ll\sigma}^{A/R}(\omega)&=&\sum_{j\neq l}  \frac{|A_{lj}|^2}{\omega+\varepsilon_j +U \langle n_{j\sigma} \rangle \mp i\delta}\,\,.
\end{eqnarray}
Equation (\ref{eqn:noise}), defining the noise spectral density in the quantum-mechanical many-body language enables a direct calculation of decoherence in various situations, as we consider next.

\section{Spectral density of noise due to Andreev fluctuators}\label{Andreevspectral}
\begin{figure}
\includegraphics[width=0.99\linewidth]{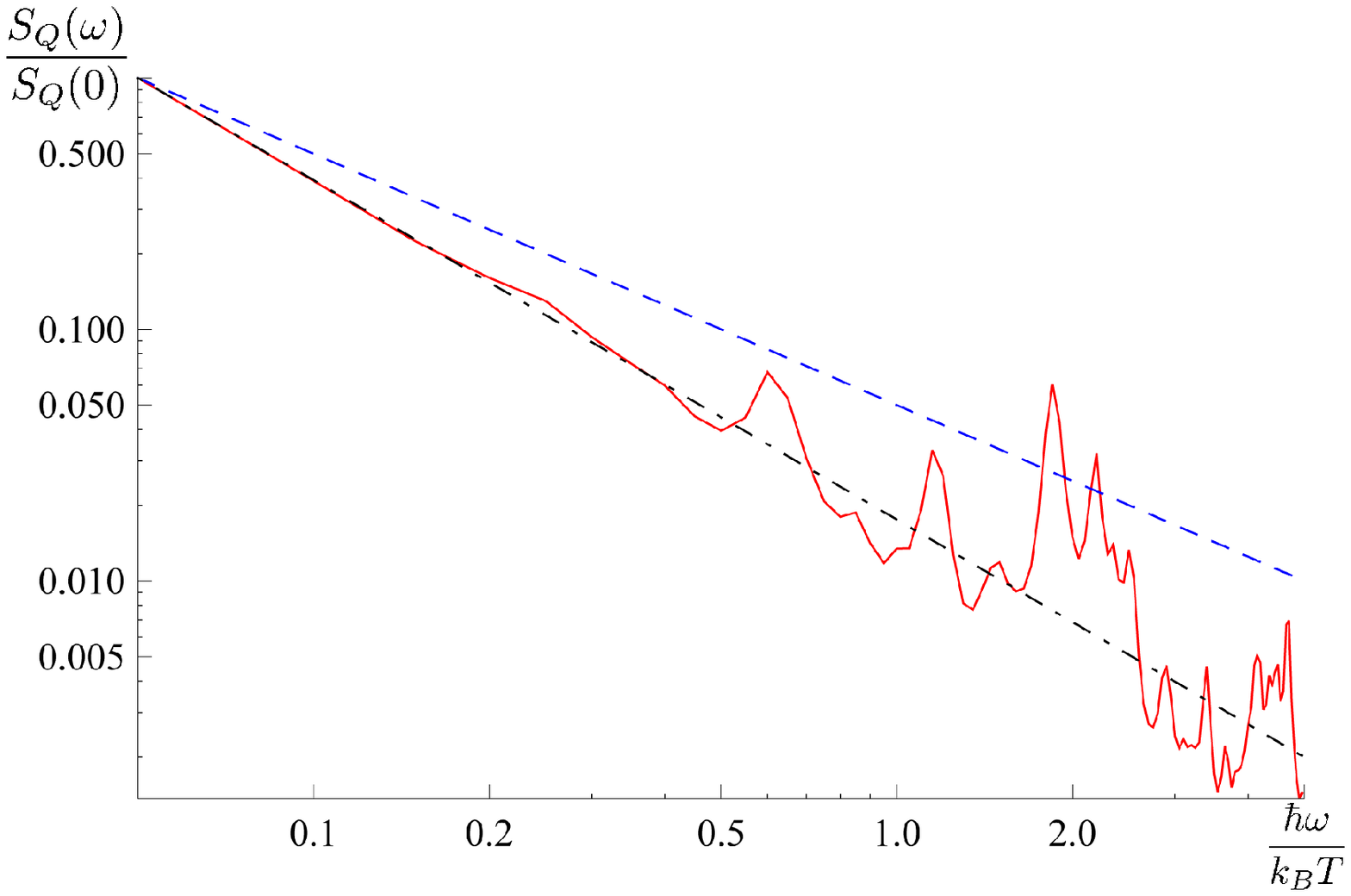}
  \caption{(color online). Log-log plot of the noise spectral density $S_Q(\omega)$ for Andreev fluctuator model.
The plot is obtained by randomly generating the positions ${\bf r}_l
\in [-5,5]\xi$ and energies $\varepsilon_l \in [-1,1]$K of 50
impurity sites and then, numerically integrating
Eq.~(\ref{eqn:noise}). Here we assumed that $v_i=v$, on-site repulsion $U \rightarrow \infty$, and the sites are occupied
with equal number of electrons with spins up and down. We used $p_F^{-1}=10^{-2}\xi$, $A_0\!=\!0.1$K and $T\!=\!0.1$K.
The dashed (blue) and dot-dashed (black) lines, shown for comparison,
correspond to $1/f$ and $1/f^{1.3}$ noise spectra. }
\label{fig:noise_lowt_Andreev}
\end{figure}

In general, the solution for $S_Q(\omega)$ with many Andreev
fluctuators, can be  obtained numerically by randomly generating the
energies $\varepsilon_l$, and positions ${\bf r}_l$ of the
impurities at the insulator/superconductor interface. The
numerically obtained spectral density of noise $S_Q(\omega)$ for 50
fluctuators is shown in Fig.~\ref{fig:noise_lowt_Andreev}. At low
frequencies the noise power spectrum has $1/f$ dependence.

For $\omega$ and $A_0$ much smaller than the typical impurity level spacing $\delta \varepsilon$ and temperature $T$, the
analytical solution for the spectral density of
noise~(\ref{eqn:noise}) is given by
\begin{align}\label{eqn:noise_small}
\!\!\!S_Q(\omega)\!\approx\!\sum_{l\sigma}4 v_l^2\! \left[1\!-\!\tanh^2\left(\frac{\tilde{\varepsilon}_{l\sigma}}{2T}\right)\right]\frac{\gamma_{l\sigma}(\tilde{\varepsilon}_{l\sigma})}{\omega^2+4\gamma_{l\sigma}(\tilde{\varepsilon}_{l\sigma})^2},
\end{align}
where $\tilde{\varepsilon}_{l\sigma}=\varepsilon_l+U\langle n_{l,-\sigma} \rangle$,
and $\gamma_{l\sigma}(\tilde{\varepsilon}_{l\sigma})=\mbox{Im}\,
\Sigma_{ll\sigma}^{A}(\tilde{\varepsilon}_{l\sigma})$ is the
broadening of the impurity energy levels due to Andreev processes.
This broadening corresponds to the fluctuations of the impurity
occupations changing the electrostatic environment of the qubit, and
thus causing dephasing. From Eq.~(\ref{eqn:noise_small}), one can
see that $S_Q(\omega)$ is given by a sum of Lorentzians with
different widths, which under proper distribution of $\gamma_l$
gives rise to a $1/f$ noise spectrum (see below). 
Given that the charge density fluctuations via Andreev processes involve two impurities with energies of the order of $T$, the probability to find two such
impurities is proportional to $(T/D)^2$ with $D$ being the impurity
energy bandwidth, and thus, $S_Q(\omega) \propto T^2 $ at low frequencies as seen experimentally.\cite{Astafiev_PRL04,Astafiev_PRL06}

For $1/f$ spectrum to arise from Eq.~(\ref{eqn:noise_small}), the distribution of $\gamma_{l}$ has to be log-normal. In order to have such distribution, the density of the charge traps has to be small, so that the dominant contribution to the self energy in Eq.~(\ref{Sigma_freq}) comes from few pairs of impurity sites, which are selected from the sum because of the energy conservation and distance constraint. Then, the switching rate $\gamma_{l} \! \propto \! \exp(-2|\mathbf{r}_{l}-\mathbf{r}_{j} |/\pi\xi)$ for a certain $j$ (see Eq.~(\ref{eqn:amplitude})). Since the distances between the charge traps are uniformly distributed, the probability of finding a switching rate $\gamma$ is $P(\gamma) \! \propto \! 1/\gamma$, leading to $1/f$ noise.
In the opposite limit of large density of charge traps, many sites $j$ contribute to the sum in Eq.~(\ref{Sigma_freq}), and the switching rates $\gamma_{l}$ self average and become approximately the same for all sites. 
Note that unlike in the theory of $1/f$ charge noise produced by fluctuating two level systems (TLS) in the substrate with log-uniform distribution in the tunnel splitting,\cite{Shnirman_PRL05} the emergence of the $1/f$ noise within Andreev fluctuator model has a qualitatively different geometrical origin due to the exponential dependence of the rate $\gamma_l$ on the distance between different impurity sites. This finding of the geometric origin of the $1/f$ noise in the Andreev fluctuator model is an important result of our work.

We note that the model of charge traps with no on-site repulsion $U=0$~[\onlinecite{Faoro_PRL05}] does not lead to $1/f$ noise because in this case the self-energy is dominated by the two-electron tunneling from the same site. The contributions to the self energy from Andreev processes involving other sites are exponentially smaller than the dominant term, and the distribution of the rates in Eq.~(\ref{eqn:noise_small}) is not log-normal. Therefore, we emphasize that the realistic model for $1/f$ noise due to Andreev processes should include both spinful fermions (to correctly describe the dynamics of charge fluctuations), and large on-site repulsion (to prevent double-electron occupation).      
 
At high frequencies $\omega \gg \delta \varepsilon, T$,  the spectral density
$S_Q(\omega)$ has resonances corresponding to the virtual processes of correlated
two-electron tunneling from (to) the impurity sites in the
insulator. These resonances, describing manifestly quantum-mechanical processes,
can be seen in Fig.~\ref{fig:noise_lowt_Andreev} at high frequencies. Their contribution to the decoherence of the qubit is suppressed by a factor $F_n(\omega t)/\omega^2$, see Eq.~(\ref{eqn:chi}). However, going beyond the pure dephasing model, $T_{1}\!\! \gg T_{2}$, considered here, one can show that correlated two-electron tunneling processes contribute to the energy relaxation of the qubit.\cite{Faoro_PRL05}

\section{The influence of pulses on decoherence} \label{sec:F}
The time dependence of the decoherence function $W_n(t)$ under a pulse sequence is given by
Eqs.~(\ref{funcint2})-(\ref{eqn:chi}), showing that the noise contribution  is modulated by a filter function
$F_{n}(\omega t)$. For the free induction decay $F_{0}(\omega t)=2\sin^2[\omega t/2]$, whereas for spin echo we have $F_{1}(\omega t)=8\sin^4[\omega t/4]$ suppressing the low-frequency ($\omega \! \ll \! 4/t$) part of $S_Q(\omega)$. In general, higher-order pulse sequences act as more efficient high-pass filters of environmental noise, i.e.~for $n$ pulses applied in time $t$ only frequencies $\omega \! > \! 2n/t$ contribute to $\chi_{n}(t)$.
Due to the formal analogy between Eq.~(\ref{eqn:chi}) and the solution for the decoherence under classical Gaussian noise, the analysis given for the latter case in Ref.~[\onlinecite{Cywinski_PRB08}] also applies here as long as the time expansion is valid. The results relevant for the noise spectral density derived here can be summarized as follows. For $S_{Q}(\omega) \propto 1/\omega^{\alpha}$, we obtain $\chi_{n}(t) \propto t^{1+\alpha}/n^{\alpha}$ for all sequences applicable to the pure dephasing case, i.e. the CPMG sequence,\cite{Haeberlen}  periodic dynamical decoupling,\cite{Viola_PRA98} concatenations of spin echo,\cite{Khodjasteh_PRL05,Khodjasteh_PRA07} and Uhrig's sequence.\cite{Uhrig_PRL07}
Thus, sequences beyond spin echo should lead to a further increase in coherence time for $1/f$ spectral density of noise.\cite{Faoro_PRL04,Falci_PRA04,Cywinski_PRB08} For noise spectrum without sharp ultra-violet cutoff, which is the case considered here, the CPMG sequence marginally outperforms other sequences.\cite{Cywinski_PRB08,Uhrig_08} Furthermore, taking into account the simplicity of CPMG sequence (defined by $\tau_{1} \! = \! \tau_{n+1} \! = \! t/2n$ and all the other $\tau_{i} \! = \! t/n$), we believe that it is a preferred approach of noise suppression for the problem at hand.~\cite{Cywinski_PRB08} We therefore propose that detailed experimental investigation of superconducting charge qubit dephasing behavior be carried out in order to test our specific predictions.

\section{Conclusion}
We consider the spin-fermion model for quantum decoherence in solid-state qubits in the pure dephasing (i.e.~$T_{1} \! \gg \! T_{2}$) situation.  We map the evolution of
the qubit interacting with the fermionic environment, possibly subject to various
$\pi$-pulse sequences, onto the Keldysh path integral. This approach is very general and allows one to apply well-developed many-body techniques to the problem of the evolution of the qubit coupled to the environment and driven by pulses. In the short-time limit, we derive the expression for the qubit decoherence which involve the product of the noise spectral density due to quantum
fluctuations of the bath and the filter function representing a particular pulse sequence. For a non-trivial interacting model of the bath,
the Andreev fluctuator model, we show that the spectral density has
$1/f$ dependence at low frequencies. Finally, we discuss the optimal strategy for the suppression of $1/f$ charge noise by the application of higher-order (beyond spin echo) pulse sequences for the problem at hand. One of our concrete conclusions of experimental significance is that the well-established CPMG pulse sequence should be an optimal method for fighting $T_{2}$ dephasing when the noise spectrum has no sharp ultra-violet cutoff.

\begin{acknowledgments}
We thank A.~Kamenev, J.~Koch,  Y.~Nakamura,
E.~Rossi, B. Shklovskii, F.~Wellstood and N.~Zimmerman for stimulating discussions.
This work was supported by the LPS-NSA-CMTC grant and by the Joint Quantum Institute (RL).
\end{acknowledgments}

%\bibliography{refs_quant}

\end{document}